\documentclass[aps,pra,showpacs,superscriptaddress,preprint]{revtex4}
\usepackage{amssymb}
\usepackage{amsmath}
\usepackage{graphicx}

\catcode`ð=\active
 \defð{\u{g}}
 \catcode`Ð=\active
 \defÐ{\u{G}}
 \catcode`Ý=\active
\defÝ{\. I}
 \catcode`ö=\active
\defö{\"{o}}
 \catcode`Ö=\active
 \defÖ{\"O}
 \catcode`ü=\active
 \defü{\"{u}}
 \catcode`Ü=\active
 \defÜ{\"{U}}
 \catcode`Þ=\active
\defÞ{\c{S}}
 \catcode`þ=\active
 \defþ{\c{s}}
 \catcode`ý=\active
 \defý{{\i}}
 \catcode`ç=\active
\defç{\d{c}}
 \catcode`Ç=\active
\defÇ{\d{C}}

\input{tcilatex}

\begin{document}

\title{Exact Klein-Gordon equation with spatially-dependent masses for
unequal scalar-vector Coulomb-like potentials }
\author{Sameer M. Ikhdair}
\email[E-mail: ]{sikhdair@neu.edu.tr;sameer@neu.edu.tr}
\affiliation{Department of Physics, Near East University, Nicosia, North Cyprus, Turkey}
\date{%
\today%
}

\begin{abstract}
We study the effect of spatially dependent mass functions over the solution
of the Klein-Gordon equation in the ($3+1$)-dimensions for spinless bosonic
particles where the mixed scalar-vector Coulomb-like field potentials and
masses are directly proportional and inversely proportional to the distance
from force center. The exact bound state energy eigenvalues and the
corresponding wave functions of the Klein-Gordon equation for mixed
scalar-vector and pure scalar Coulomb-like field potentials are obtained by
means of the Nikiforov-Uvarov (NU) method. The energy spectrum is discussed
for different scalar-vector potential mixing cases and also for constant
mass case.

\end{abstract}

\maketitle

\newpage

\section{Introduction}

The relativistic wave equations can be considered as first approximation to
the field theory when the corrections are taken in the presence of strong
potential fields [1-3]. This explains the increased interest in the
Klein-Gordon (KG) and Dirac wave equations to find exact and analytic
solutions for energy spectrum and wavefunctions. As a consequence of the
physical importance of exact solutions of relativistic KG wave equation in
quantum mechanics, under the influence of strong potentials, an increasing
interest in this equation has appeared in the study of quark-antiquark mass
spectroscopy [4-6], atomic, nuclear, and plasma physics [7-9]. So, the idea
is to use the bosonic particle KG equation as a mathematical tool to reach
the goal of obtaining approximate solutions for nonrelativistic fermionic
particles eigenstates. The problems that can be solved exactly in
relativistic quantum mechanics are very limited perhaps because of the
mathematical difficulties.

Recently, the bound and scattering solutions of the $s$- and $l$-waves KG
and Dirac equations for any interaction system have raised a great interest
[10-12]. The bound-states of the Dirac and KG equations with the
Coulomb-like scalar plus vector potentials have been studied in arbitrary
dimension [13-17]. Furthermore, the exact results for the scattering states
of the KG equation with Coulomb-like scalar plus vector potentials have been
investigated in an arbitrary dimension [18].\ This equation has been exactly
solved for a larger class of linear, exponential and linear plus Coulomb
potentials to determine the bound state energy spectrum using two
semiclassical methods [19]. Many authors have considered a more general
transformation between the unequal vector and scalar potentials given by%
\begin{equation}
V(r)=V_{0}+\beta S(r),
\end{equation}%
where $V_{0}$ and $\beta $ being arbitrary constants of certain proportions
have to be chosen after solving the problem under consideration [19-21]. It
is interesting to note that, this restriction includes the case where $%
V(r)=0,$ when both constants vanish, the situation where the potentials are
equal in magnitude and sign $V(r)=S(r)$ or equal in magnitude but opposite
in sign $V(r)=-S(r)$ (\textit{i.e}., $V_{0}=0;\beta =\pm 1),$ and also the
case where the potentials are proportional when $V_{0}=0$ [21]$.$ For the
case when $S(r)\geq V(r),$ there exist bound state solutions. However, it
should be noticed that the case when the scalar potential is equal to the
vector potential must be considered separately [3]. Under the condition of $%
S(r)=V(r),$ the KG turns into a Schr\"{o}dinger-like equation and thus the
bound state solutions are very easily to obtain with the help of the
well-known methods developed in the non-relativistic quantum mechanics.

On the other hand, the problem of the spatially-dependent effective mass is
presenting a growing interest along the last few years [22-25]. Many authors
have used different methods to study the partially exactly solvable and
exactly solvable Schr\"{o}dinger, KG and Dirac equations in the presence of
variable mass having a suitable mass distribution functions in $1D,3D$
and/or any $D$-dimensional cases for different potentials, such as the
linear potential [20], the exponential-type potentials [21], the Coulomb
potential [26], the Lorentz scalar interactions [27], the hyperbolic-type
potentials [28], the Morse potential [29], the P\"{o}schl-Teller potential
[30], the inversely linear scalar potential [31], the Coulomb and harmonic
potentials [32], the modified Kratzer-type, rotationally corrected Morse
potentials [33], Mie-type and pseudoharmonic potentials [34]. Recently, the
point canonical transformation (PCT) has also been employed to solve the $D$%
-dimensional spatially dependent mass Schr\"{o}dinger equation for some
molecular potentials to get the exact bound state solutions including the
energy spectrum and corresponding wave functions [32-34]. It is quite
natural to look for relativistic treatment of this type of systems, mostly
because the ordering ambiguity which is present in the nonrelativistic case
[35], is expected to be avoided under relativistic ambiance [36,37].

Very recently, the NU method has been used to solve any $l$-states KG
equation approximately for the Hulth\'{e}n potential with a suitable choice
of spatially-dependent mass function distribution of an exponential-type
[38]. Also, a new approximation scheme [39] has been proposed for the
centrifugal term to obtain a quasi-exact analytic bound-state solution of
the radial KG equation with spatially-dependent effective mass for scalar
and vector Hulth\'{e}n potentials in any arbitrary dimension and orbital
angular momentum quantum number $l$ within the framework of the NU method
[40].

The problem of a particle subject to an inversely linear potential in one
spatial dimension ($\sim \left\vert x\right\vert ^{-1}$), known as the
one-dimensional hydrogen atom, by considering a convenient mixing of vector
and scalar Lorentz structures has received considerable attention in the
literature [41,42]. The same problem for a spinless particle subject to a
general mixing of vector and scalar inversely linear potentials in the (1$+1$%
)-dimensional world was analyzed [43]. Exact bounded solutions were found in
closed form by imposing boundary conditions on the eigenfunctions which
ensure that the effective Hamiltonian is Hermitian for all the points of the
space. Jia and Souza Dutra [44] considered position-dependent effective mass
Dirac equations with $PT$ and non-$PT$ symmetric potentials. Souza Dutra and
Jia [31], investigated the exact solution of the (1$+1$)-dimensional KG
equation with spatially dependent mass for the inversely linear potential.
Recently, in ref. [45] the bound state solutions of the (1$+1$)-dimensional
KG equation with mass inversely proportional to the distance from the force
center for the inversely linear potential were obtained by using the NU
method [46-52]. Two particular cases are studied, the case when vector
potential is equal to the scalar potential in magnitude $V(r)=S(r)$ and when
vector potential is equal to the scalar potential in magnitude but not in
sign $V(r)=-S(r),$ (\textit{i.e.}, $V_{0}=0;\beta =\pm 1).$

In the present work, we feel tempted to extend the work of ref. [45] to
study the bound state solutions of the ($3+1$)-dimensional KG equation with
position-dependent bosonic mass function $m(r)=m_{0}\left( 1+\lambda
_{0}br^{-1}\right) $ where $r\neq 0$ [45] for the attractive scalar
potential $S(r)=-\kappa _{s}r^{-1}$ with $\kappa _{s}=\hbar cq_{s}$ being
the coupling constant, taking into consideration the general mixings of
scalar and vector Lorentz structure potential given in eq. (1)$.$ Firstly,
this choice of mass function together with potential mixings is mostly
suitable for modeling some physical systems like the Kratzer-type potentials
[48]. Secondly, the motivation for this choice is due to the nature of the
dominating Coulombic field between the two interacting nuclei at short
distances. Thirdly, this choice enables one to solve the KG equation easily
and elegantly. The general mixings of potentials include: (i) $V(r)=-\kappa
_{v}r^{-1}$ and $S(r)=0,$ which represents a $\pi ^{-\text{ }}$ meson in a
Coulomb field. (ii) $V(r)=0$ and $S(r)=-\kappa _{s}r^{-1},$ which has no
experimental evidence. (iii) $V(r)=$ $S(r)=-\kappa r^{-1},$ $(\kappa
_{s}=\kappa _{v}=\kappa ,$ $V_{0}=0,$ $\beta =1)$ which represents not only
a KG particle in an equally mixed Lorentz scalar and Lorentz vector
potentials but also a Dirac particle in the same potential mixture, where $%
l=j+1/2$ and the radial KG wave function represents the radial
large-component of the Dirac spinor [53]. (iv) $V(r)=-\kappa _{v}r^{-1}+Ar$
and $S(r)=0$ representing a $\pi ^{-\text{ }}$ meson in a Coulomb field
perturbed by a linear Lorentz vector interaction $Ar.$ Also, we consider the
effect of a spatially dependent mass of the linear form $m(r)=m_{0}r/L$ [45]
on the solution of the ($3+1$)-dimensional KG equation for the Lorentz
vector and scalar potentials of the form $V(r)=0$ and $S(r)=sr^{-1},$
respectively. It is worth mentioning that this choice of mass function
together with potential mixings is mostly suitable for modeling some
physical systems like the Pseudoharmonic potential.

The paper is organized as follows. In sect. 2, we outline the NU method.
Section 3 is devoted for the bound state analytic solutions of the ($3+1$%
)-dimensional KG equation with spatially dependent mass functions for two
quantum systems obtained by means of the NU method. Finally, the relevant
results are discussed in sect. 4.

\section{NU Method}

The NU method is briefly outlined here and the details can be found in ref.
[46]. This method is proposed to solve the second-order differential
equation of the hypergeometric type: 
\begin{equation}
\psi _{n}^{\prime \prime }(z)+\frac{\widetilde{\tau }(z)}{\sigma (z)}\psi
_{n}^{\prime }(z)+\frac{\widetilde{\sigma }(z)}{\sigma ^{2}(z)}\psi
_{n}(z)=0,
\end{equation}%
where $\sigma (z)$ and $\widetilde{\sigma }(z)$ are polynomials, at most, of
second-degree, and $\widetilde{\tau }(s)$ is a first-degree polynomial. In
order to find a particular solution for eq. (2), let us decompose the
wavefunction $\psi _{n}(z)$ as follows:%
\begin{equation}
\psi _{n}(z)=\phi _{n}(z)y_{n}(z),
\end{equation}%
and use%
\begin{equation}
\left[ \sigma (z)\rho (z)\right] ^{\prime }=\tau (z)\rho (z),
\end{equation}%
to reduce eq. (2) to the form%
\begin{equation}
\sigma (z)y_{n}^{\prime \prime }(z)+\tau (z)y_{n}^{\prime }(z)+\lambda
y_{n}(z)=0,
\end{equation}%
with%
\begin{equation}
\tau (z)=\widetilde{\tau }(z)+2\pi (z),\text{ }\tau ^{\prime }(z)<0,
\end{equation}%
where the prime denotes the differentiation with respect to $z.$ One is
looking for a family of solutions corresponding to%
\begin{equation}
\lambda =\lambda _{n}=-n\tau ^{\prime }(z)-\frac{1}{2}n\left( n-1\right)
\sigma ^{\prime \prime }(z),\ \ \ n=0,1,2,\cdots ,
\end{equation}%
The $y_{n}(z)$ can be expressed in terms of the Rodrigues relation:%
\begin{equation}
y_{n}(z)=\frac{B_{n}}{\rho (z)}\frac{d^{n}}{dz^{n}}\left[ \sigma ^{n}(z)\rho
(z)\right] ,
\end{equation}%
where $B_{n}$ is the normalization constant and the weight function $\rho
(z) $ is the solution of the differential equation (4). The other part of
the wavefunction (3) must satisfy the following logarithmic equation%
\begin{equation}
\frac{\phi ^{\prime }(z)}{\phi (z)}=\frac{\pi (z)}{\sigma (z)}.
\end{equation}%
By defining 
\begin{equation}
k=\lambda -\pi ^{\prime }(z).
\end{equation}%
one obtains the polynomial

\begin{equation}
\pi (z)=\frac{1}{2}\left[ \sigma ^{\prime }(z)-\widetilde{\tau }(z)\right]
\pm \sqrt{\frac{1}{4}\left[ \sigma ^{\prime }(z)-\widetilde{\tau }(z)\right]
^{2}-\widetilde{\sigma }(z)+k\sigma (z)},
\end{equation}%
where $\pi (z)$ is a parameter at most of order $1.$ The expression under
the square root sign in the above equation can be arranged as a polynomial
of second order where its discriminant is zero. In this regard, an equation
for $k$ is being obtained. After solving such an equation, the $k$ values
are determined through the NU method.

\section{Exact Bound-State Solutions}

In the relativistic quantum mechanics, for a spinless particle, we write the
full stationary KG equation for a spatially dependent bosonic mass in real ($%
3+1$)-dimensions as [49,50] 
\begin{equation}
\mathbf{\nabla }^{2}\psi (\mathbf{r})+\frac{1}{\hbar ^{2}c^{2}}\left\{ \left[
E_{nl}-V(r)\right] ^{2}-\left[ m(r)c^{2}+S(r)\right] ^{2}\right\} \psi (%
\mathbf{r})=0,\text{ }\nabla ^{2}=\sum\limits_{j=1}^{3}\frac{\partial ^{2}}{%
\partial x_{j}^{2}}
\end{equation}%
where $m(r)$ is a bosonic mass, $E_{nl}$ is the energy of the particle, $%
V(r) $ is a Lorentz vector (coupled as the $0$-component of the four-vector
potential) and $S(r)$ is a Lorentz scalar (added to the mass term)
potentials. Let us decompose the radial wavefunction $\psi (\mathbf{r})$ as
follows: 
\begin{equation}
\psi (\mathbf{r})=\frac{u(r)}{r}Y_{m}^{(l)}(\widehat{\mathbf{r}}),
\end{equation}%
where $u(r)$ is the radial wave function and $Y_{m}^{(l)}(\widehat{\mathbf{r}%
})$ is the angular dependent spherical harmonics and this reduces eq. (12)
into the following position-dependent effective mass Schr\"{o}dinger-like
equation:%
\begin{equation*}
\frac{d^{2}u(r)}{dr^{2}}+\frac{1}{\hbar ^{2}c^{2}}\left\{
E_{nl}^{2}+V(r)^{2}-2EV(r)-m(r)^{2}c^{4}-S(r)^{2}-2m(r)c^{2}S(r)-\frac{%
l(l+1)\hbar ^{2}c^{2}}{r^{2}}\right\}
\end{equation*}%
\begin{equation}
\times u(r)=0,\text{ }u(0)=0.
\end{equation}%
Now, we will start to analyze some illustrative particular cases for Lorentz
scalar-vector mixings and suitable mass distribution functions for modeling
some important physical systems.

\subsection{Mixed vector-scalar Coulomb potentials}

We shall present a general solution of ref. [46] for general admixture of
scalar and vector potential mixings. Let us solve eq. (14) for the general
relationship between vector and scalar potentials given in eq. (1), then we
have 
\begin{equation*}
-\frac{d^{2}u(r)}{dr^{2}}+\frac{1}{\hbar ^{2}c^{2}}\left[ \left( 1-\beta
^{2}\right) S(r)^{2}-2\left( \left( E_{nl}+V_{0}\right) \beta +m(r)\right)
S(r)+\frac{l(l+1)\hbar ^{2}c^{2}}{r^{2}}\right] u(r)
\end{equation*}%
\begin{equation}
=\frac{1}{\hbar ^{2}c^{2}}\left[ \left( E_{nl}+V_{0}\right)
^{2}-m^{2}(r)c^{4}\right] u(r).
\end{equation}%
Furthermore, we take the scalar potential in the form of an attractive
Coulomb-like field 
\begin{equation}
S(r)=-\frac{\hbar cq_{s}}{r},\text{ }q_{s}=q,\text{ }r\neq 0
\end{equation}%
where $q_{s}$ is being a scalar dimensionless real parameter coupling
constant and $\hbar c$ is being a constant with $J.fm$ dimension. At this
stage, it is worthwhile to mention that the above choice of the mass
function of Coulombic form together with the presently taken admixture of
scalar and vector fields are mostly suitable for modeling the well-known
pseudo-Coulomb (Kratzer-type) potentials [48]. Following refs. [38,40,45],
we may take the spatially-dependent mass function being described by%
\begin{equation}
m(r)=m_{0}\left( 1+\frac{\lambda _{0}b}{r}\right) ,\text{ }r\neq 0,
\end{equation}%
where $m_{0}$ and $\lambda _{0}=\frac{\hbar }{m_{0}c}$ are the rest mass of
the bosonic particle and the Compton-like wavelength in $fm$ units,
respectively; $b$ is a dimensionless real constant. The interaction field
has much impact on the choice of the mass function which, in the present
case, is inveresely proportional to the distance between the two nuclei at
short distances $m(r)\sim \frac{1}{r}$ and constant at long distances $%
m(r\rightarrow \infty )\simeq m_{0}$. Let us introduce the variable change $%
z=r$ $\in (0,\infty )$ and define%
\begin{equation*}
\varepsilon _{nl}=\frac{\sqrt{m_{0}^{2}c^{4}-\widetilde{E}_{nl}^{2}}}{Q}%
\text{ \ }(m_{0}c^{2}\geq \widetilde{E}_{nl}),\text{ }\gamma _{1}=\frac{%
2\left( b-q\right) m_{0}c^{2}-2q\beta \widetilde{E}_{nl}}{Q},\text{ }
\end{equation*}%
\begin{equation}
\widetilde{E}_{nl}=E_{nl}+V_{0},\text{ }\gamma _{2}=b\left( b-2q\right)
+q^{2}\left( 1-\beta ^{2}\right) +l\left( l+1\right) ,\text{ }Q=\hbar c,
\end{equation}%
with the following constraint $V_{0}\leq -E_{nl}+m_{0}c^{2}$ must be
fulfilled for bound state solutions. Further, substituting eqs. (16)-(18)
into eq. (15), we obtain 
\begin{equation}
\frac{d^{2}u(z)}{dz^{2}}-\left( \frac{\varepsilon _{nl}^{2}z^{2}+\gamma
_{1}z+\gamma _{2}}{z^{2}}\right) u(z)=0,\text{ }\unit{u}(0)=0,
\end{equation}%
where $u(z)=u(r)$. In the present work, we deal with bound state solutions,
the quantum condition is obtained from the finiteness of the solution at
infinity, \textit{i.e}., the wave function $u(r)$ must satisfy boundary
conditions, $u(r)=0$ when $r\rightarrow \infty $ and at the origin point, $%
r=0.$ In order to solve eq. (19) by means of the NU method, we should
compare it with eq. (2). The following values for parameters are found:%
\begin{equation}
\widetilde{\tau }(z)=0,\text{\ }\sigma (z)=z,\text{\ }\widetilde{\sigma }%
(z)=-\left( \varepsilon _{nl}^{2}z^{2}+\gamma _{1}z+\gamma _{2}\right) .
\end{equation}%
Inserting these values of parameters into eq. (11), we obtain 
\begin{equation}
\pi (z)=\frac{1}{2}\pm \frac{1}{2}\sqrt{4\varepsilon _{nl}^{2}z^{2}+4(\gamma
_{1}+k)z+4\gamma _{2}+1}.
\end{equation}%
The discriminant of the square root must be set equal to zero, \textit{i.e}%
., $\Delta =4\varepsilon _{nl}^{2}z^{2}+4(\gamma _{1}+k)z+4\gamma _{2}+1=0.$
Consequently, the following two constants $k_{1}$ and $k_{2}$ are found to be%
\begin{equation}
k_{1,2}=-\gamma _{1}\pm \varepsilon _{nl}\sqrt{1+4\gamma _{2}},
\end{equation}%
with the following requirements on the parameters $\beta \leq 2\sqrt{(l+%
\frac{1}{2})^{2}+(q-b)^{2}}$ and $V_{0}\leq -E_{nl}+m_{0}c^{2}$ must be
fulfilled for real solutions. In this regard, we can find the possible
functions for $\pi (z)$ as 
\begin{equation}
\pi (z)=\left\{ 
\begin{array}{cc}
\frac{1}{2}\pm \left[ \varepsilon _{nl}z+\frac{1}{2}\sqrt{1+4\gamma _{2}}%
\right] & \text{\ for }k_{1}=-\gamma _{1}+\varepsilon _{nl}\sqrt{1+4\gamma
_{2}}, \\ 
\frac{1}{2}\pm \left[ \varepsilon _{nl}z-\frac{1}{2}\sqrt{1+4\gamma _{2}}%
\right] & \text{\ for }k_{2}=-\gamma _{1}-\varepsilon _{nl}\sqrt{1+4\gamma
_{2}}.%
\end{array}%
\right.
\end{equation}%
According to the NU method, one of the four values of the polynomial $\pi
(z) $ is just proper to obtain the energy states because $\tau (z)$ has a
negative derivative for this value of $\pi (z).$ Therefore, the selected
forms of $\pi (z)$ and $k$ take the following particular values%
\begin{equation}
\pi (z)=-\varepsilon _{nl}z+\frac{1}{2}\left( 1+\sqrt{1+4\gamma _{2}}\right)
,\text{ }k=-\gamma _{1}-\varepsilon _{nl}\sqrt{1+4\gamma _{2}},
\end{equation}%
to obtain%
\begin{equation}
\tau (z)=-2\varepsilon _{nl}z+1+\sqrt{1+4\gamma _{2}},\text{ }\tau ^{\prime
}(z)=-2\varepsilon _{nl}<0,
\end{equation}%
where $\tau ^{\prime }(z)=\frac{d\tau (z)}{dz}.$ In addition, after using
eqs. (24) and (25) together with the assignments given in eq. (20), the
following expressions for $\lambda $ are obtained%
\begin{equation}
\lambda _{n}=\lambda =2n\varepsilon _{nl},\text{ }n=0,1,2,\cdots ,
\end{equation}%
\begin{equation}
\lambda =-\gamma _{1}-\varepsilon _{nl}\left( 1+\sqrt{1+4\gamma _{2}}\right)
.
\end{equation}%
Letting $\lambda _{n}=\lambda ,$ we can solve the above equations for the
energy states $E_{nl}^{\pm }$ as

\begin{equation}
E_{nl}^{\pm }=-V_{0}+\frac{\left[ q\left( b-q\right) \beta \pm B_{nl}\sqrt{%
B_{nl}^{2}-q^{2}(1-\beta ^{2})-b\left( b-2q\right) }\right] }{q^{2}\beta
^{2}+B_{nl}^{2}}m_{0}c^{2},
\end{equation}%
where%
\begin{equation}
B_{nl}=n+\frac{1}{2}+\sqrt{\left( l+\frac{1}{2}\right) ^{2}+b\left(
b-2q\right) +q^{2}\left( 1-\beta ^{2}\right) },\text{ }n,l=0,1,2,\cdots .
\end{equation}%
For spatially-dependent mass case, \textit{i.e.}, $b\neq 0,$ the bound state
solutions of the system are determined by the parameters $q$ and $b.$ It is
not difficult to conclude that all bound-states appear in pairs, two energy
solutions are valid for the particle $E^{p}=E_{nl}^{+}$ and the second one
corresponds to the anti-particle energy $E^{a}=E_{nl}^{-}$ in the
Coulomb-like field.

Let us now find the corresponding eigenfunctions for this system. Using eqs.
(4) and (9), we find%
\begin{equation}
\rho (z)=z^{\sqrt{1+4\gamma _{2}}}e^{-2\varepsilon _{nl}z},
\end{equation}%
\begin{equation}
\phi (z)=z^{\left( 1+\sqrt{1+4\gamma _{2}}\right) /2}e^{-\varepsilon _{nl}z}.
\end{equation}%
Hence, substituting eq. (30) into eq. (8), we find%
\begin{equation}
y_{n}(z)=D_{n}z^{-\sqrt{1+4\gamma _{2}}}e^{2\varepsilon _{nl}z}\frac{d^{n}}{%
dz^{n}}\left[ z^{\left( n+\sqrt{1+4\gamma _{2}}\right) }e^{-2\varepsilon
_{nl}z}\right] \sim L_{n}^{2L+1}(2\varepsilon _{nl}z),
\end{equation}%
where $L_{n}^{\alpha }(x)$ is the generalized Laguerre polynomials. By using 
$u(z)=\phi (z)y_{n}(z),$ we get the wavefunctions as%
\begin{equation}
u(r)=Nr^{\left( 1+\sqrt{1+4\gamma _{2}}\right) /2}e^{-\varepsilon
_{nl}r}L_{n}^{2L+1}(2\varepsilon _{nl}r),
\end{equation}%
where%
\begin{equation}
L=\sqrt{\left( l+\frac{1}{2}\right) ^{2}+b\left( b-2q\right) +q^{2}\left(
1-\beta ^{2}\right) }-\frac{1}{2}.
\end{equation}%
Using the normalization condition $\int_{0}^{\infty }u(r)^{2}dr=1$ and the
orthogonality relation of the generalized Laguerre polynomials $%
\int_{0}^{\infty }x^{\alpha +1}e^{-x}\left[ L_{n}^{(\alpha )}(x)\right]
^{2}dx=\left( 2n+\alpha +1\right) \frac{\Gamma (n+\alpha +1)}{n!}$, the
normalizing factor $N$ can be found as [54-57]%
\begin{equation}
N=\sqrt{\frac{n!\left( 2\varepsilon _{nl}\right) ^{2L+3}}{2(n+L+1)\Gamma
(n+2L+2)}},
\end{equation}%
where $\varepsilon _{nl}$ and $L$ are given in eqs. (18) and (34),
respectively.

(1) If we consider the case when scalar potential is equal the vector
potential in magnitude and sign,\textit{\ i.e.}, $V_{0}=0$ and $\beta =1,$
then we have 
\begin{subequations}
\begin{equation}
E_{nl}^{p}=\frac{\left[ q\left( b-q\right) +B_{nl}\sqrt{B_{nl}^{2}-b\left(
b-2q\right) }\right] }{q^{2}+B_{nl}^{2}}m_{0}c^{2},
\end{equation}%
\begin{equation}
\text{ }E_{nl}^{a}=\frac{\left[ q\left( b-q\right) -B_{nl}\sqrt{%
B_{nl}^{2}-b\left( b-2q\right) }\right] }{q^{2}+B_{nl}^{2}}m_{0}c^{2},
\end{equation}%
where 
\end{subequations}
\begin{equation}
B_{nl}=n+\frac{1}{2}+\sqrt{\left( l+\frac{1}{2}\right) ^{2}+b\left(
b-2q\right) },\text{ }n,l=0,1,2,\cdots .
\end{equation}%
Obviously, the bound state solutions of the particle and anti-particle are
available. When the mass is taken to be constant, i.e., $b=0,$ we have%
\begin{equation}
E_{nl}^{p}=\frac{\left( n+l+1\right) ^{2}-q^{2}}{\left( n+l+1\right)
^{2}+q^{2}}m_{0}c^{2},\text{ }E_{nl}^{a}=-m_{0}c^{2},\text{ }%
n,l=0,1,2,\cdots ,
\end{equation}%
where $n$ and $l$ signify the usual radial and orbital quantum numbers. The
particle has bound state solution whereas anti-particle has continuum
solution for all states.

(2) If we consider the case when scalar potential is equal to the vector
potential in magnitude but not in sign, i.e., $V_{0}=0$ and $\beta =-1,$
then we have 
\begin{subequations}
\begin{equation}
E_{nl}^{p}=\frac{\left[ -q\left( b-q\right) +B_{nl}\sqrt{B_{nl}^{2}-b\left(
b-2q\right) }\right] }{q^{2}+B_{nl}^{2}}m_{0}c^{2},
\end{equation}%
\begin{equation}
E_{nl}^{a}\text{ }=\frac{\left[ -q\left( b-q\right) -B_{nl}\sqrt{%
B_{nl}^{2}-b\left( b-2q\right) }\right] }{q^{2}+B_{nl}^{2}}m_{0}c^{2}.
\end{equation}%
For the constant-mass case, \textit{i.e.}, $b=0,$ we have 
\end{subequations}
\begin{equation}
E_{nl}^{p}=m_{0}c^{2},\text{ }E_{nl}^{a}=-\frac{\left( n+l+1\right)
^{2}-q^{2}}{\left( n+l+1\right) ^{2}+q^{2}}m_{0}c^{2}.
\end{equation}%
Obviously, the particle has continuum solution for all states whereas bound
state solution for anti-particle. In addition, when the potential coupling
constant is taken as $q=b/2,$ the spectra of the varying mass KG particle in
potential fields $q_{s}=q_{v}$ ($q_{s}=-q_{v}$) are similar to the spectra
of constant mass KG particle in the potential fields $q_{s}=-q_{v}$ ($%
q_{s}=q_{v}$), respectively.

\subsection{Pure Scalar Coulomb-like potential}

In their paper [44], Souza Dutra and Jia used the pure scalar potential that
is inversely proportional to the absolute value of the coordinate. Here, we
use a pure scalar repulsive Coulomb-like field potential%
\begin{equation}
S(r)=\frac{s}{r},\text{ }V(r)=0,
\end{equation}%
with $s$ being a coupling parameter with $J.fm$ dimension and also assume
the spatially-dependent mass function having a linear form%
\begin{equation}
m(r)=Ar,
\end{equation}%
with $A=\frac{m_{0}}{L}$ where $m_{0}$ is being the rest mass and $L$ is
being a constant with space dimension. At this stage, it is worthwhile to
mention that the above choice of the mass function together with the
presently taken pure scalar potential case are mostly suitable for modeling
the well-known pseudoharmonic potentials [58,59]. Inserting eqs. (41) and
(42) into eq. (15), then we have%
\begin{equation}
-\frac{d^{2}u(r)}{dr^{2}}+\frac{1}{\hbar ^{2}c^{2}}\left[ \frac{%
m_{0}^{2}c^{4}}{L^{2}}r^{2}+\frac{s^{2}+l(l+1)\hbar ^{2}c^{2}}{r^{2}}\right]
u(r)=\frac{1}{\hbar ^{2}c^{2}}\left( E_{nl}^{2}-\frac{2m_{0}c^{2}s}{L}%
\right) u(r).
\end{equation}%
Thus, the present problem has been reduced to three-dimensional Schr\"{o}%
dinger equation for pseudoharmonic oscillator problem which was solved
before in refs. [58,59]. Introducing the variable change $z=r^{2}$ $\in
(0,\infty )$ and defining%
\begin{equation}
\varepsilon _{nl}=\frac{1}{\hbar c}\sqrt{\frac{2m_{0}c^{2}s}{L}-E_{nl}^{2}}%
,\ \alpha _{1}=\frac{m_{0}c}{\hbar L},\text{ }\alpha _{2}=\frac{%
s^{2}+l(l+1)\hbar ^{2}c^{2}}{\hbar ^{2}c^{2}},
\end{equation}%
we obtain 
\begin{equation}
\frac{d^{2}u(z)}{dz^{2}}+\frac{1}{2z}\frac{du(z)}{dz}+\frac{1}{\left(
2z\right) ^{2}}\left( -\alpha _{1}^{2}z^{2}-\varepsilon _{nl}^{2}z-\alpha
_{2}\right) u(z)=0,
\end{equation}%
where $u(z)=u(r).$ Comparing eq. (45) with eq. (2),\ we find values for the
parameters as%
\begin{equation}
\widetilde{\tau }(z)=1,\text{\ }\sigma (z)=2z,\text{\ }\widetilde{\sigma }%
(z)=-\alpha _{1}^{2}z^{2}-\varepsilon _{nl}^{2}z-\alpha _{2},
\end{equation}%
and by inserting these values of parameters into eq. (11), we further obtain 
\begin{equation}
\pi (z)=\frac{1}{2}\pm \frac{1}{2}\sqrt{4\alpha _{1}^{2}z^{2}+4(\varepsilon
_{nl}^{2}+k)z+4\alpha _{2}+1},
\end{equation}%
and the constant $k$ as%
\begin{equation}
k_{1,2}=-\varepsilon _{nl}^{2}\pm \alpha _{1}\sqrt{4\alpha _{2}+1}.
\end{equation}%
When the individual values of $k$ given in eq. (48) are being substituted
into Eq. (47), the four possible forms of $\pi (z)$ are written as follows%
\begin{equation}
\pi (z)=\left\{ 
\begin{array}{cc}
\frac{1}{2}\pm \frac{1}{2}\left[ 2\alpha _{1}z+\sqrt{4\alpha _{2}+1}\right]
& \text{\ for }k_{1}=-\varepsilon _{nl}^{2}+\alpha _{1}\sqrt{4\alpha _{2}+1},
\\ 
\frac{1}{2}\pm \frac{1}{2}\left[ 2\alpha _{1}z-\sqrt{4\alpha _{2}+1}\right]
& \text{\ for }k_{2}=-\varepsilon _{nl}^{2}-\alpha _{1}\sqrt{4\alpha _{2}+1}.%
\end{array}%
\right.
\end{equation}%
According to the NU method, the selected forms of $\pi (z)$ and $k$ are
taking the following particular values%
\begin{equation}
\pi (z)=-\alpha _{1}z+\frac{1}{2}\left[ 1+\sqrt{4\alpha _{2}+1}\right] ,%
\text{ }k=-\varepsilon _{nl}^{2}-\alpha _{1}\sqrt{4\alpha _{2}+1},
\end{equation}%
to obtain%
\begin{equation}
\tau (z)=-2\alpha _{1}z+2+\sqrt{4\alpha _{2}+1},\text{ }\tau ^{\prime
}(z)=-2\alpha _{1}<0,
\end{equation}%
which is the essential condition in the method. Also, the following
expressions for $\lambda $ are obtained%
\begin{equation}
\lambda _{n}=\lambda =2n\alpha _{1},\text{ }n=0,1,2,\cdots ,
\end{equation}%
\begin{equation}
\lambda =-\varepsilon _{nl}^{2}-\alpha _{1}\left( 1+\sqrt{4\alpha _{2}+1}%
\right) .
\end{equation}%
Letting $\lambda _{n}=\lambda ,$ we can solve the above equations for the
energy eigenvalues $E_{nl}$ as%
\begin{equation}
\widetilde{E}_{nl}=\alpha _{1}\left( 1-2A_{nl}\right) ,
\end{equation}%
where%
\begin{equation}
\widetilde{E}_{nl}=-\varepsilon _{nl}^{2},\text{ }A_{nl}=-\left( n+\frac{1}{2%
}\sqrt{\left( 2l+1\right) ^{2}+\frac{4s^{2}}{\hbar ^{2}c^{2}}}\right) ,
\end{equation}%
and thus we find%
\begin{equation}
\frac{E_{nl}^{2}}{m_{0}c^{2}}=\frac{2s}{L}+\frac{\hbar c}{L}\left( 2n+1+%
\sqrt{\left( 2l+1\right) ^{2}+\frac{4s^{2}}{\hbar ^{2}c^{2}}}\right) ,
\end{equation}%
which is found to be consistent with eq. (20) of ref. [51] obtained by SUSY
method when $l$ is set equal to zero. We note that the energy levels for
particles and antiparticles are symmetric about $E_{nl}=0$ [49]$.$

Essentially, we should report that eq. (43) corresponds to the Schr\"{o}%
dinger equation of anharmonic oscillator potential $V(r)=\alpha
_{1}^{2}r^{2},$ with energy levels [58,59]%
\begin{equation}
\widetilde{E}_{nl}=\alpha _{1}(2n+2\Lambda +3),
\end{equation}%
with $\widetilde{E}_{nl}$ is given in (55), $\alpha _{1}$ in (44) and $%
\Lambda $ is defined by%
\begin{equation}
\Lambda =\frac{1}{2}\left( \sqrt{\left( 2l+1\right) ^{2}+\left( \frac{2s}{%
\hbar c}\right) ^{2}}-1\right) .
\end{equation}%
Let us now find the corresponding eigenfunctions for this system. After
using eqs. (4) and (9), we find%
\begin{equation}
\rho (z)=z^{\sqrt{4\alpha _{2}+1}/2}e^{-\alpha _{1}z},
\end{equation}%
\begin{equation}
\phi (z)=z^{\left( 1+\sqrt{4\alpha _{2}+1}\right) /4}e^{-\alpha _{1}z/2}.
\end{equation}%
Substituting eq. (59) into eq. (8), we obtain%
\begin{equation}
y_{n}(z)=D_{n}z^{-\sqrt{4\alpha _{2}+1}/2}e^{\alpha _{1}z}\frac{d^{n}}{dz^{n}%
}\left[ z^{\left( n+\sqrt{4\alpha _{2}+1}/2\right) }e^{-\alpha _{1}z}\right]
\sim L_{n}^{\left( 2\Lambda +1\right) /2}(\alpha _{1}z).
\end{equation}%
By using $u(z)=\phi (z)y_{n}(z),$ we get the wavefunctions as%
\begin{equation}
u(r)=Ne^{-\alpha _{1}r^{2}/2}r^{\left( \Lambda +1\right) /2}L_{n}^{\left(
2\Lambda +1\right) /2}(\alpha _{1}r^{2}),
\end{equation}%
where $\Lambda $ is defined in eq. (58). It is worth mentioning that the
above wave function is consistent with eq. (20) of ref. [58] in the solution
of the Schr\"{o}dinger equation for the pseudoharmonic oscillator potential.
Essentially, such a solution has been discussed before by many authors [59].
Making use of the normalization condition $\int_{0}^{\infty }u(r)^{2}dr=1$
and the orthogonality relation of the generalized Laguerre polynomials $%
\int_{0}^{\infty }x^{\alpha ^{\prime }-1}e^{-x}\left[ L_{n}^{(\alpha )}(x)%
\right] ^{2}dx=\left( 
\begin{array}{c}
\alpha -\alpha ^{\prime }+n \\ 
n%
\end{array}%
\right) \Gamma (\alpha ^{\prime })$, the normalization constant $N$ can be
found as [54-57]%
\begin{equation}
N=\sqrt{\frac{2\left( \frac{m_{0}c}{\hbar L}\right) ^{\frac{1}{2}\sqrt{%
\left( 2l+1\right) ^{2}+\left( \frac{2s}{\hbar c}\right) ^{2}}+1}}{\left( 
\begin{array}{c}
n-1 \\ 
n%
\end{array}%
\right) \Gamma \left( \frac{1}{2}\sqrt{\left( 2l+1\right) ^{2}+\left( \frac{%
2s}{\hbar c}\right) ^{2}}+1\right) }}.
\end{equation}

\section{Conclusions}

We have extended the (1$+1$)-dimensional KG solution in ref. [45] to the $l$%
-waves KG for scalar-vector mixing Coulomb-like fields with suitable choices
of spatially dependent mass functions. Thus, for this kind of studied
problems, we may conclude that the relativistic wave equation can be solved
exactly. For suitable choices of potential forms as the general mixing of
scalar-vector and pure scalar Coulomb-like field potential, the relativistic
bound state energy spectrum and wave functions have been obtained,
respectively. The resulting solutions of the wave functions are being
expressed in terms of the generalized Laguerre polynomials. We have
considered different mass functions of inversely proportional and directly
proportional to the coordinate distance. Obviously, when the coupling
potential parameters are adjusted to some specific values, particularily
when $q=b/2,$ the spectra of the mass varying KG particle for the case $%
q_{s}=q_{v}$ ($q_{s}=-q_{v}$) become similar to the spectra of the constant
mass KG particle for the case $q_{s}=-q_{v}$ ($q_{s}=q_{v}$), respectively.
It is found that the KG equation with a suitable mass function for a pure
scalar potential is being reduced into the constant mass Schr\"{o}dinger
equation for the anharmonic oscillator potential. In the limit of constant
mass ($b=0$), the solution for the energy eigenvalues and wave functions are
reduced to those ones in literature. Also, when $l=0,$ the problem reduces
to $s$-waves solution.

\acknowledgments The author thanks the anonymous kind referee for the
positive and invaluable suggestions that have improved the manuscript
greatly. He is also grateful for the partial support provided by the
Scientific and Technological Research Council of Turkey (T\"{U}B\.{I}TAK).

\newpage

{\normalsize 
}

\end{document}